# Robust multichannel bulk transport in a time-reversal-invariant insulator-free photonic waveguide array

Guochao Wei[1,*], Yingfeng Qi[2,*], Kang Du[1,†], Wei Zhu[1], Zhenzhen Liu[3], Junjun Xiao[4,‡], Shengxiang Wang[1,5 §], and Zhen Gao[2,¶],

[1]School of Microelectronics, Wuhan Textile University, Wuhan 430200, China

[2]State Key Laboratory of Optical Fiber and Cable Manufacturing Technology, Department of Electronic and Electrical Engineering,

Guangdong Key Laboratory of Integrated Optoelectronics Intellisense, Southern University of Science and Technology, Shenzhen 518055, China

[3]Research Center for Advanced Optics and Photoelectronics, Department of Physics, College of Science, Shantou University, Shantou 515063, China

[4]School of Integrated Circuits, Harbin Institute of Technology (Shenzhen), Shenzhen 518055, China

[5]State Key Laboratory of New Textile Materials and Advanced Processing Technologies, Wuhan Textile University, Wuhan 430200, China

∗ These authors contributed equally to this work.

† dk@wtu.edu.cn (K. D.); ‡ eiexiao@hit.edu.cn (J. X.); § shxwang@wtu.edu.cn (S. W.); ¶gaoz@sustech.edu.cn (Z. G.)

**High-density photonic integration requires closely packed channels that combine low crosstalk with robust transport. However, achieving both in an insulator-free photonic architecture while preserving time-reversal symmetry remains challenging. Here, we theoretically propose and experimentally demonstrate a time-reversal-invariant, insulator-free photonic waveguide array that supports robust multichannel bulk transport with low crosstalk. The design exploits boundary-induced chiral anomaly bulk states (CABSs) in two types of dielectric photonic crystals with Dirac points at the Brillouin-zone center (Γ) and corners (K/K′), respectively. The large momentum mismatch between these states suppresses interchannel coupling even when neighboring channels are in direct contact. Microwave experiments confirm low-crosstalk multichannel bulk transport and robustness against metallic obstacles and voids. We further demonstrate independently addressable cavity modes in directly adjacent triangular resonators and low-crosstalk transmission through a coplanar waveguide crossing with a tailored intersection. These results establish a nonmagnetic design strategy that combines efficient spatial utilization, momentum-selective isolation, and defect-tolerant bulk transport, enabling compact photonic architectures beyond parallel waveguide arrays.**

## Introduction

The pursuit of ultracompact and efficient photonic systems has driven the development of strategies to suppress crosstalk [1–9] and enhance robustness against perturbations [10–13]. In conventional waveguide arrays, reducing the channel separation increases evanescent coupling between neighboring waveguides, resulting in unwanted crosstalk and signal degradation at high integration densities [14–21]. Confinement strategies based on artificial gauge fields [22–24] and meta-grating claddings [25–27] can mitigate this coupling, but the required cladding regions limit spatial utilization. More recently, dispersion-engineered, cladding-free waveguide arrays have emerged as an alternative platform for high-density photonic integration [28]. As schematically shown in Fig. 1(a), by deliberately shifting the equi-frequency contours of neighboring channels, these arrays introduce a mismatch between their wavevectors, strongly suppressing inter-channel coupling even at zero channel separation [28–30]. Each channel thus serves as both a transmission medium and an effective cladding for its neighbors, enabling near-complete spatial utilization and low crosstalk. Nevertheless, these architectures predominantly rely on topologically trivial bulk modes. Although momentum mismatch provides channel isolation, it does not, by itself, confer topological robustness against sharp bends, defects, or structural disorder.

Topological photonics offers a promising route to combining dense integration with robust transport [31–40]. Unlike conventional topological edge states, which are localized at interfaces, topological bulk modes can occupy the available cross-sectional area of a finite structure, thereby enabling efficient spatial utilization [41–43]. In particular, boundary-induced chiral anomaly bulk states (CABSs) can emerge in finite Dirac systems when boundary terminations selectively admit bulk modes compatible with the imposed electromagnetic boundary conditions [44–46]. These states extend throughout the bulk and support robust transport within finite-size-induced pseudogaps. However, previous demonstrations of topological bulk transport have generally relied on additional cladding regions for mode confinement and inter-channel isolation, limiting spatial utilization. Recent experiments have demonstrated insulator-free topological photonic multilane highways in gyromagnetic photonic crystals with broken time-reversal symmetry, enabling robust, cladding-free multichannel chiral valley bulk transport [47], as schematically shown in Fig. 1(b). Although these architectures offer higher integration densities and greater information capacity, their reliance on gyromagnetic materials and external magnetic fields makes extension to optical frequencies challenging. This raises a central question: can a multichannel photonic waveguide array simultaneously achieve high spatial utilization, low crosstalk, and robust transport while preserving time-reversal symmetry?

Here, we theoretically propose and experimentally demonstrate a time-reversal-invariant, insulator-free multichannel photonic waveguide array that combines boundary-induced bulk transport with momentum-selective channel isolation, as schematically shown in Fig. 1(c). Two types of Dirac photonic crystals, with Dirac points at Γ and K/K′, respectively, are directly joined along their zero-flux symmetry planes to generate CABSs with distinct wavevectors. The large momentum mismatch between these modes strongly suppresses interchannel coupling even at zero channel separation. Microwave experiments demonstrate low-crosstalk multichannel bulk transport and robustness against various perturbations such as metallic obstacles and air defects. We further extend this mechanism to two-dimensional routing architectures, experimentally demonstrating direct-adjacent triangular resonators and a coplanar waveguide crossing with negligible mutual crosstalk. Our results establish a nonmagnetic route toward densely integrated topological photonic circuits that combine efficient spatial utilization, low crosstalk, and robust bulk transport.

**Results**

We begin by considering a 2D honeycomb PC (PC1) whose unit cell comprises six dielectric cylinders (red dots) [48]. Owing to Brillouin zone (BZ) folding, a fourfold-degenerate Dirac point emerges at the center of the BZ, as shown in Fig. 2(a). The presence of mirror symmetry along the $x$-axis (green dashed lines in Fig. 2(b)) allows the eigenstates to be classified according to their parity as even or odd modes. For the even modes (modes 1 and 3), the energy flux (black arrows) attains its maximum along the mirror plane (green dashed line), whereas for the odd modes (modes 2 and 4), the energy flux vanishes on the mirror plane. Based on this configuration, we further consider a photonic waveguide constructed from PC1, as shown in the right panel of Fig. 2(c). The waveguide is bounded by PECs at $y = 0$ and $y = L$, placed along symmetry-compatible (energy flux vanished) planes parallel to the propagation direction. The Hamiltonian of the waveguide with thickness $L$ can be expressed as

$$H(y) = v_D\left(k_x\tau_z\sigma_x - i\partial_y\sigma_y\right) + m\sigma_z \quad (1)$$

where $v_D$ denotes the group velocity, $\sigma_i(i = x, y, z)$ are Pauli matrices acting on the orbital (pseudospin) basis, $\tau_z = \pm 1$ labels the two valley-like subspaces that are time-reversal partners, and $m$ is an effective mass term. Because the structure is terminated by PEC boundaries, the electromagnetic boundary conditions impose a fixed phase relationship among the field components at the boundaries, thereby inducing a spatially dependent mass term in the vicinity of the Dirac cone, $m(y) = m_1,\ y < 0$; $m(y) = 0, 0 \le y \le L$; $m(y) = m_2,\ y > L$. In the domain of $0 \le y \le L$ (PC1 region), the bulk Hamiltonian has a mirror-$y$ symmetry, $\sigma_x H(y)\sigma_x = H(-y)$. Hence, the eigensolutions at the Dirac point in each valley subspace can be obtained as [44]

$$\psi = \begin{cases} e^{|m_1|y/v_D}\begin{pmatrix}1\\ \mathrm{sgn}(m_1)\end{pmatrix}, & y < 0 \\ \begin{pmatrix}1\\ \pm 1\end{pmatrix}, & 0 \le y \le L \\ e^{-|m_2|(y-L)/v_D}\begin{pmatrix}1\\ -\mathrm{sgn}(m_2)\end{pmatrix}, & y > L \end{cases} \quad (2)$$

The introduction of PEC boundaries can be viewed as imposing spatially dependent mass terms in the effective Hamiltonian, with $m_1$ for $y < 0$ and $m_2$ for $y > L$. The continuity of the wave function at $y = 0$ and $L$ constrains the signs of $m_1$ and $m_2$: $m_1 > 0,\ m_2 < 0$ for even bulk modes, and $m_1 < 0,\ m_2 > 0$ for odd bulk modes. Because the tangential electric field is required to vanish at the PEC boundaries, only the two odd bulk modes (modes 2 and 4) in the vicinity of the Dirac point satisfy the imposed boundary conditions. As a consequence, two CABSs emerge in a pseudogap (orange region) originating from the finite-size effect associated with the PEC-induced mass inversion (see Section 1 of the Supplemental Material for detailed theoretical analysis of their formation [49]), as shown by the red lines in Fig. 2(c). The corresponding eigenfield profiles, shown in the right panels of Fig. 2(c), exhibit spatially extended distributions across the entire bulk region of the waveguide.

To realize a Dirac point at the corner of the BZ, we consider a triangular PC (denoted as PC2) whose unit cell consists of a single dielectric cylinder (blue dot), as shown in the inset of Fig. 2(d). The simulated bulk band structure in Fig. 2(d) exhibits a Dirac point at the K point, protected by $C_{3v}$ symmetry [50]. The two eigenstates near the Dirac point (green and yellow dots) are shown in Fig. 2(e), displaying even (mode 1) and odd (mode 2) eigenfield profiles with respect to the $x$-axis (green dashed line), respectively. When PC2 is truncated by two PEC boundaries [right panel of Fig. 2(f)], only the odd mode satisfies the boundary condition (varnished energy flux), giving rise to a single CABS near K and K′ valleys (see the detailed properties and layer-dependent band dispersions of the CABSs in Sections 2 and 3 of the Supplemental Material [49]), as shown by the blue lines in Fig. 2(f). Directly joining PC1 and PC2 at the symmetry-compatible truncation planes yields CABS-derived modes predominantly confined to their respective channels. The large wavevector mismatch between these modes strongly suppresses inter-channel mode conversion, enabling effectively decoupled bulk transport without an intervening cladding region.

To experimentally demonstrate this CABS-based insulator-free photonic waveguide array, we construct a composite structure by directly joining PC1 and PC2 at their truncated centers (green dashed lines in Fig. 2(b) and 2(e)). Figure 3(a) presents a photograph of the fabricated sample, which consists of dielectric cylinders embedded in an air foam substrate and enclosed between metallic plates. The red and blue circles and half-circles in the inset denote the constituent cylinders of PC1 and PC2, respectively, while the green dashed line marks their direct interface. The upper metallic plate has been removed to see the internal structure. At the same operating frequency, the PC1- and PC2-derived modes exhibit distinct interface-parallel Bloch wavevectors in the common projected BZ. This wavevector mismatch suppresses phase-matched inter-channel power transfer, enabling low-crosstalk propagation without an intervening cladding region. Figure 3(b) shows the simulated band dispersion of the low-crosstalk and cladding-free photonic waveguide array. The gray lines denote the trivial bulk modes, whereas the red and blue curves represent the CABS associated predominantly with the upper and lower channels, respectively. The corresponding eigenfield distributions in the right panel show that each CABS remains largely confined to its own channel, indicating limited inter-channel hybridization and strongly suppressed crosstalk. To experimentally demonstrate the wave propagation characteristics of this CABS-based cladding-free waveguide array, we use a point source (green star) to excite and a probe to measure the electric-field distributions of the two CABSs, as shown in Figs. 3(c) and 3(e). The excited waves are predominantly localized within the upper [Fig. 3(c)] and lower [Fig. 3(e)] channels, respectively, with negligible leakage into the adjacent channel. This strong confinement originates from the large momentum mismatch between the CABSs in the two neighboring channels. To quantitatively evaluate the isolation between the two neighboring channels, we further measured their transmission spectra, as shown in Figs. 3(d) and 3(f). When the point source is placed in the upper channel, the transmission measured in the same channel (red line) is significantly higher than that in the lower channel (blue line), clearly demonstrating low crosstalk between the adjacent two channels. Conversely, when the point source is placed in the lower channel, the transmission there (blue line) greatly exceeds that in the upper channel (red line). The simulated field distributions and the electric-field-based crosstalk analysis are provided in Section 4 of the Supplemental Material. We note that these CABS-based, insulator-free, and low-crosstalk photonic waveguide arrays can be readily extended to systems with multiple channels (see Section 5 of the Supplemental Material for details [49]).

Next, we investigate the robustness of the time-reversal-invariant, insulator-free photonic waveguide array against two representative perturbations: a metallic obstacle, indicated by the yellow bars in Figs. 4(a) and 4(b), and a void created by removing four unit cells of dielectric cylinders, marked by the gray squares in Figs. 4(c) and 4(d). Each perturbation is introduced separately into the upper and lower channels. The electric-field distributions measured at 10.7 GHz are shown in Figs. 4(a) and 4(c). For both perturbation types, the CABSs redistribute around the defect and subsequently recover their original propagation profiles, with no discernible standing-wave fringes or backscattering. More importantly, the fields remain predominantly confined to the initially excited channel, while the neighboring channel remains nearly unexcited, indicating that neither perturbation induces significant inter-channel coupling or observable crosstalk. The corresponding simulated field distributions at 11 GHz, shown in Figs. 4(b) and 4(d), reproduce the experimentally observed defect bypassing, suppressed backscattering, and strong channel selectivity, further supporting the robustness of the CABSs against both metallic obstacles and air voids. Robust transport with low crosstalk is also maintained around sharp bends, as detailed in Supplemental Section 6 [49]. We attribute the observed tolerance to perturbations primarily to the spatially extended mode profiles of the CABSs. Because their modal energy is distributed across several lattice rows, a localized perturbation overlaps with only a fraction of that energy, allowing the field to redistribute around the perturbed region and recover its original modal profile downstream. This reduced fractional overlap may offer an advantage over narrowly confined interface modes under comparable perturbations, as a defect located at the field maximum of a narrowly confined mode can interact with a larger fraction of its modal energy. A detailed comparison with a narrowly confined topological interface state is provided in Supplemental Section 7 [49]. We further demonstrate directly adjacent triangular resonators and a low-crosstalk coplanar waveguide crossing without additional isolation regions, as detailed in the End Matter.

## Conclusion

In summary, we have theoretically proposed and experimentally demonstrated a time-reversal-invariant, insulator-free multichannel photonic waveguide array based on momentum-separated CABSs. The large mismatch between the projected wavevectors of CABSs in directly adjoining photonic crystals suppresses inter-channel mode conversion, enabling independently addressable bulk transport with low crosstalk. Their spatially extended bulk field profiles facilitate robust transport around metallic obstacles, air defects, and sharp bends. We have further extended this approach to two-dimensional architectures, realizing nested triangular resonators that share a common boundary and a low-crosstalk coplanar waveguide crossing without additional isolation regions. These results establish boundary-induced CABSs as a foundation for topological photonic circuits that combine dense integration, low crosstalk, and robust bulk transport while preserving time-reversal symmetry. The proposed strategy may be extended to higher-frequency regimes and to acoustic and mechanical wave systems. Integration with nonlinear or gain media could further enable compact platforms for exploring nonlinear and active topological phenomena.

## Acknowledgements

This work was supported by National Natural Science Foundation of China (12504510, 62375064, 62361166627, 62375118 and 52402153), National Key R&D Program of China (2025YFA1412300), Natural Science Foundation of Hubei Province (2026AFB673), Guangdong Basic and Applied Basic Research Foundation (2024A1515012770), Shenzhen Science and Technology Program (ZDCY20250901100959001, JCYJ20250604145307010), Shenzhen Science and Technology Innovation Commission (202308073000209), and High-level Special Funds

(G03034K004), Natural Science Foundation of Wuhan (2025040601020165), State Key Laboratory of New Textile Materials and Advanced Processing, Wuhan Textile University (FZ2025007), and Hubei Provincial Engineering Research Center for Wide-Bandgap Semiconductor Materials and Devices Open Project (2025WBGMD04).

**Data and materials availability:** All data are available in the main text and the supplemental information.

**Competing interests:** The authors declare no competing interests.

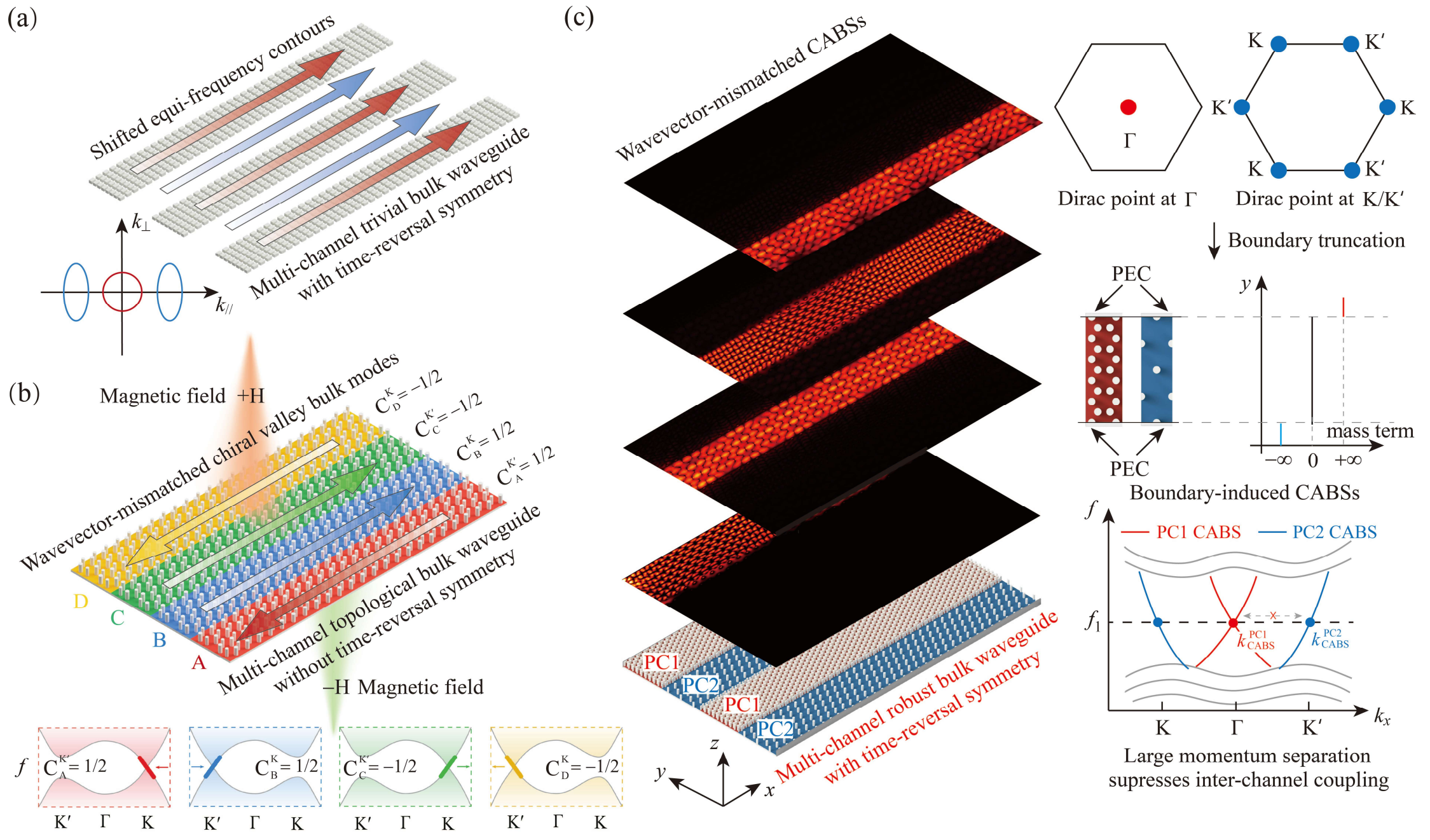

FIG. 1. Design principle of time-reversal-invariant, insulator-free multichannel photonic waveguide arrays. (a) Conventional cladding-free waveguide array with shifted equifrequency contours (EFCs). The separation between the EFCs of neighboring channels introduces a wavevector mismatch that suppresses interchannel coupling, enabling high-density transport with low crosstalk. In this configuration, the time-reversal symmetry is preserved, and the multichannel waveguide array is topologically trivial. (b) Insulator-free topological photonic multilane highway constructed from photonic valley half-semimetal domains arranged in an A–B–C–D sequence. The A–B–C and B–C–D heterostructures support K′- and K-valley-locked chiral bulk modes. Valley-momentum mismatch suppresses interchannel coupling, but this architecture relies on a magnetic bias and broken time-reversal symmetry. (c) Time-reversal-invariant, insulator-free multichannel photonic waveguide array comprising alternating PC1 and PC2. PC1 hosts a Dirac point at Γ, whereas PC2 hosts time-reversal-related Dirac points at K and K′. Within an effective Dirac model, PEC terminations at symmetry-compatible truncation planes are represented by infinite-mass regions of opposite signs on either side of a massless strip, supporting boundary-induced chiral anomaly bulk states (CABSs) within finite-size-induced pseudogaps. Upon direct joining of PC1 and PC2, the resulting modes retain the characteristic dispersions and bulk-extended profiles of the isolated CABSs, while their large projected-wavevector mismatch suppresses interchannel conversion. The stacked-field distributions illustrate independently addressable, multi-channel bulk transport with low inter-channel crosstalk at zero channel separation, under time-reversal symmetry.

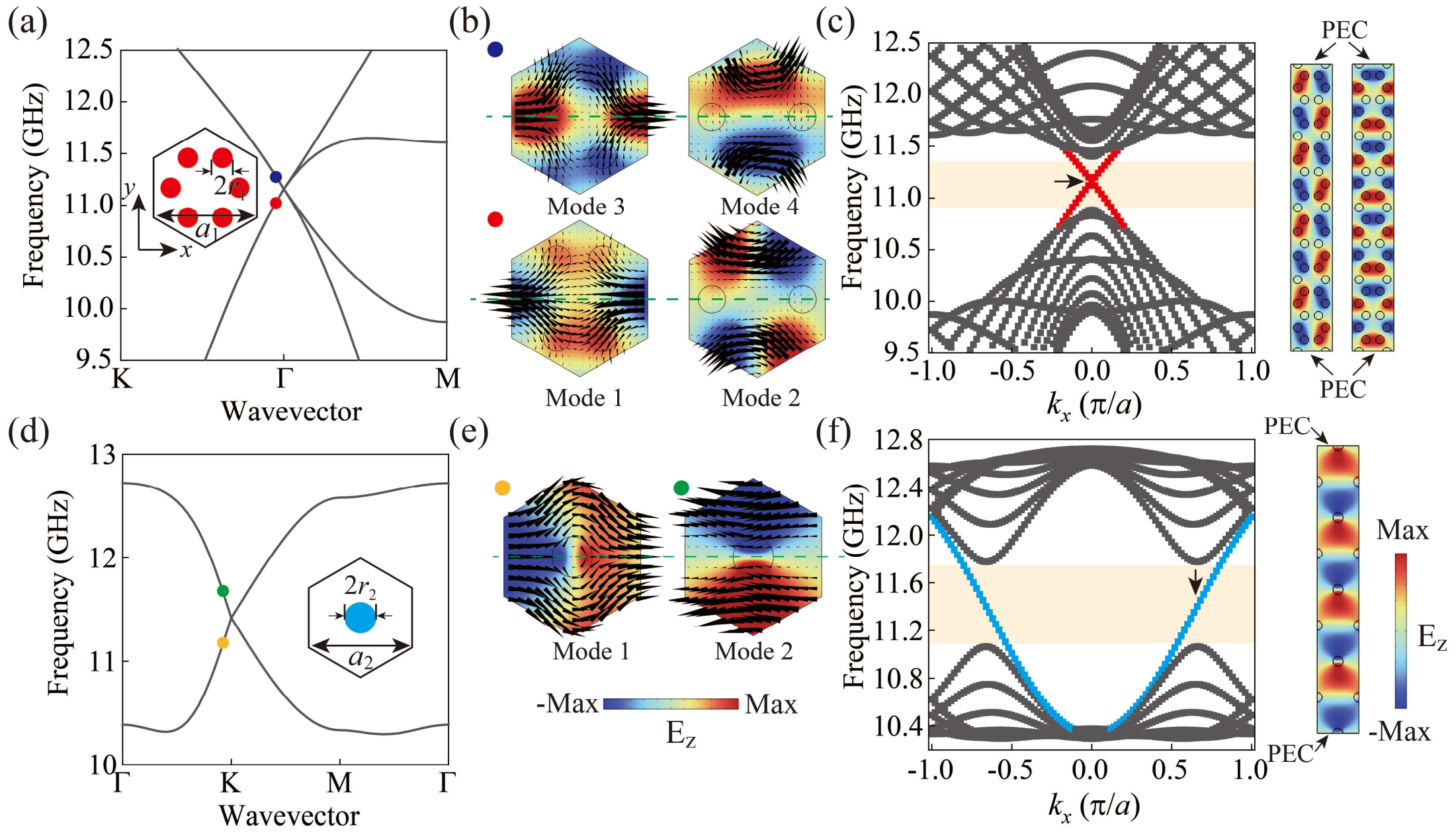

FIG. 2. Boundary-induced CABSs in two different Dirac photonic crystals. (a) Simulated bulk band structure of honeycomb PC (PC1), featuring a Dirac cone located at the center (Γ point) of the BZ. Inset shows the unit cell of PC1, which consists of six dielectric cylinders with radius $r_1 = 1.5$ mm, lattice constant $a_1 = 15$ mm, and permittivity $\varepsilon_1 = 9.8$. (b) Simulated $E_z$ field distributions (colormap) and energy fluxes (black arrows) of the eigenmodes near the Dirac point (red and blue dots in (a)). (c) Simulated band dispersion of PC1 truncated by two PEC boundaries, where the gray and red lines denote the bulk modes and boundary-induced CABSs, respectively. Right panel: Eigenfield distributions of the boundary-induced CABSs. (d) Simulated bulk band structure of a triangular PC (PC2), featuring a Dirac cone located at the corner (K point) of the BZ. The inset shows the unit cell of PC2 consists of a single dielectric cylinder with $a_2 = 15$ mm, $r_2 = 1.8$ mm, and $\varepsilon_2 = 24$. (e) Simulated $E_z$ field distributions (colormap) and energy fluxes (black arrows) of the eigenmodes near the Dirac point (yellow and green dots in (d)). (f) Simulated band dispersion of PC2 truncated by two PEC boundaries, where the gray and blue lines denote the bulk modes and boundary-induced CABSs, respectively. Right panel: eigenfield distributions of the CABSs.

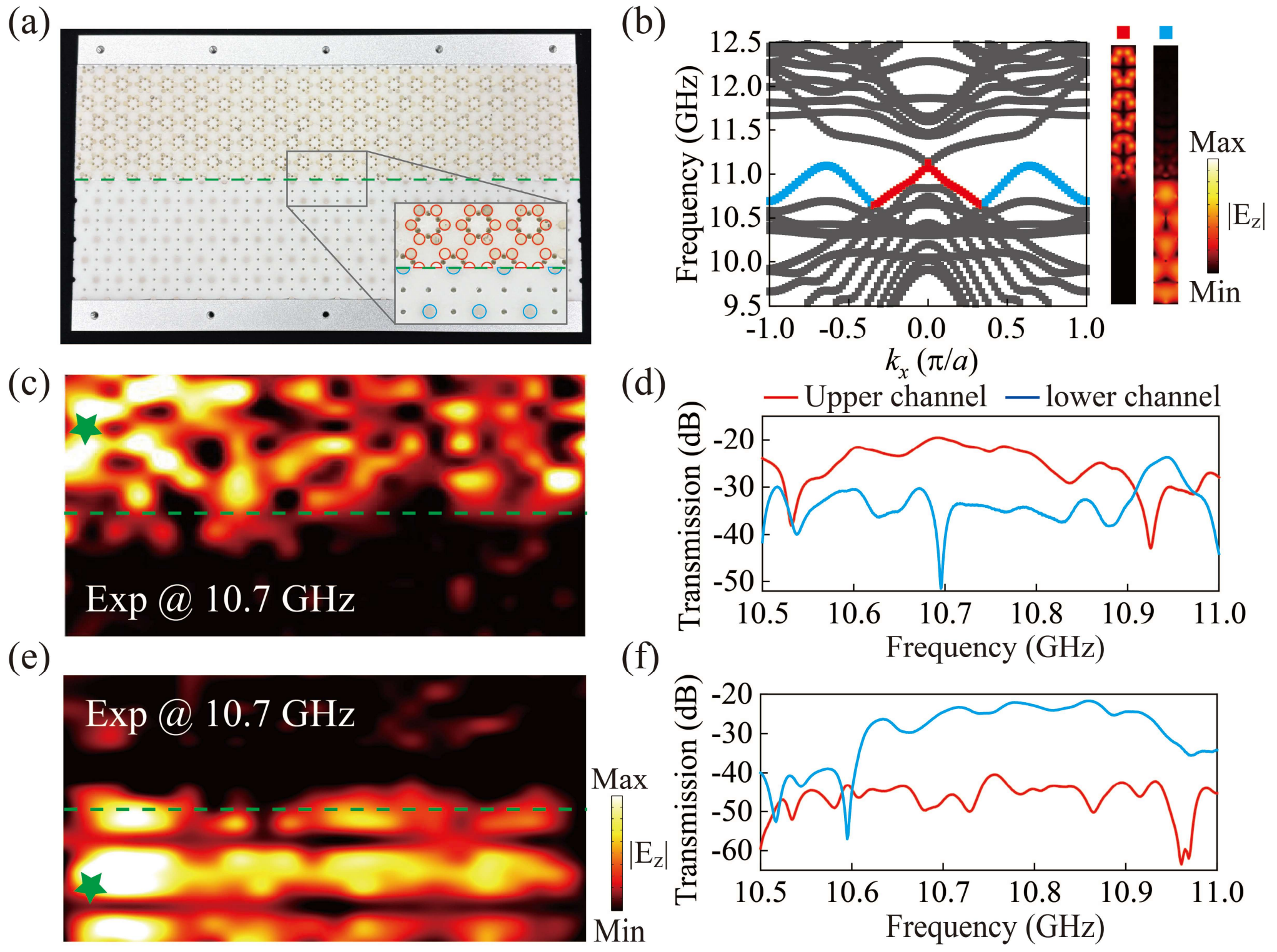

FIG. 3. Experimental realization of a robust, low-crosstalk, and insulator-free photonic multichannel bulk waveguide array based on CABSs in PCs. (a) Photograph of the fabricated sample. Dielectric cylinders are embedded in air foam, forming a honeycomb lattice (PC1) in the upper region and a triangular lattice (PC2) in the lower region. The inset shows a magnified view of the interface (green dashed line) between the two PCs. (b) Simulated band dispersion of the cladding-free waveguide array formed by PC1 and PC2. Gray lines denote bulk modes, while red and blue lines represent the low-crosstalk CABSs in the upper and lower channels, respectively. The right panel shows the corresponding eigenfield profiles. (c) Measured electric-field distribution and (d) transmission spectra of the low-crosstalk CABSs under upper-channel excitation (green star). Red and blue curves correspond to the transmission spectra of the upper and lower channels, respectively. (e) and (f) Same as (c) and (d), respectively, but for lower-channel excitation.

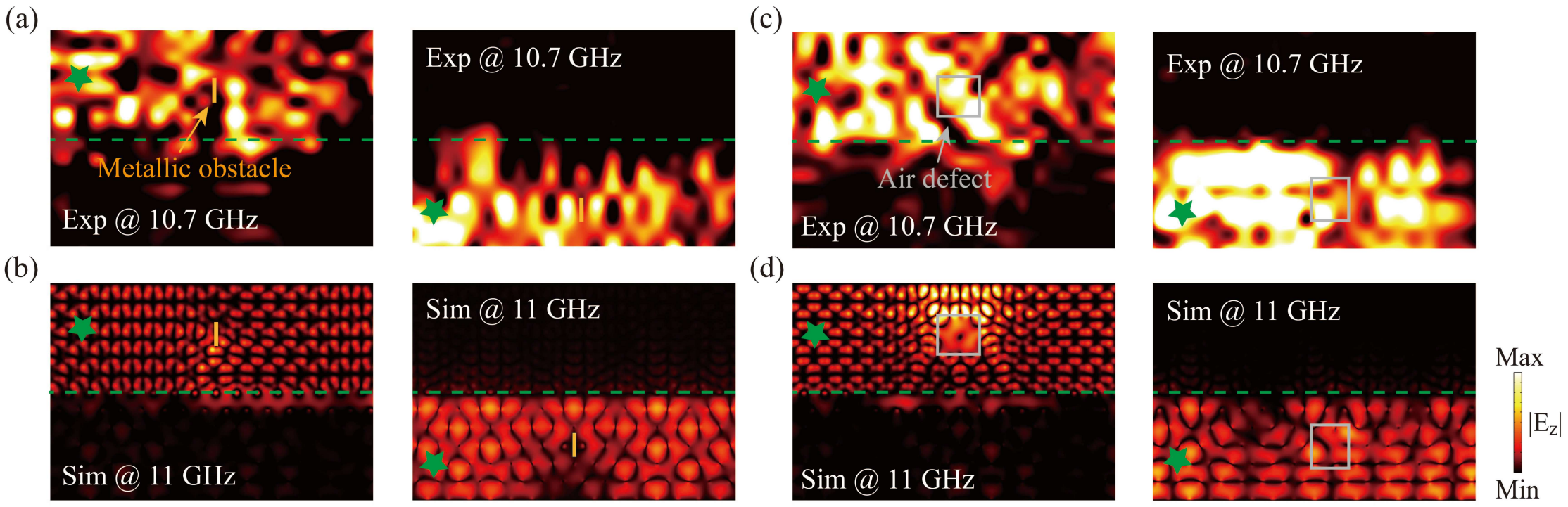

FIG. 4. Robustness of the insulator-free, low-crosstalk photonic waveguide array based on CABSs. (a) Measured and (b) simulated electric-field distributions for a metallic obstacle (orange bar) placed in the upper (left) and lower (right) channels. (c) and (d) Same as (a) and (b) but with an air defect (gray square).

## End Matter

Beyond enabling robust, low-crosstalk propagation in one dimension, the concept of CABS-based, cladding-free photonic waveguide arrays can be extended to two-dimensional geometries to realize robust, cladding-free resonators. Figure 5(a) shows a photograph of the nested triangular resonator structure, in which an inner domain consisting of four layers of PC1 is directly surrounded by an outer domain comprising six layers of PC2. The green dashed triangle marks the closed boundary shared by the two domains, with no additional cladding or spatial separation. The electric-field distributions measured at 10.7 GHz are shown in Figs. 5(b) and 5(c). When the outer resonator mode is excited, the field remains predominantly confined to the PC2 domain, with negligible penetration into the adjacent inner domain [Fig. 5(b)]. Conversely, excitation of the inner resonator mode produces strong field localization within the PC1 domain, while the outer domain remains nearly unexcited [Fig. 5(c)]. These results demonstrate that directly adjoining resonators can support independently addressable resonant modes with limited mutual leakage, suppressing inter-resonator crosstalk without conventional cladding or physical spacing. Moreover, the fields follow the triangular paths around all three sharp corners without appreciable scattering or leakage, highlighting the robustness of closed-loop CABS transport. The simulated field distributions at 11 GHz, shown in Figs. 5(d) and 5(e), reproduce the experimentally observed mode localization and spatial profiles. The small frequency offset between the experimental and simulated results is primarily attributed to fabrication tolerances and the finite air gap in the experimental configuration.

Beyond parallel multichannel waveguide arrays, the momentum selectivity of CABSs can be exploited to realize planar waveguide crossings. At a crossing, the overlapping channels locally alter the photonic crystal environment, potentially opening additional scattering pathways that cause reflection or mode conversion. To mitigate these effects, we introduce a tailored intersection region, denoted PC3, as illustrated in Fig. 6(a). The Γ- and K/K′-associated CABSs occupy well-separated regions of momentum space and possess distinct projected Bloch wavevectors. Accordingly, PC3 is designed to support Dirac degeneracies at Γ and K/K′, providing states with compatible momenta to facilitate transmission through the intersection while suppressing conversion between the crossing channels. The calculated band structure in Fig. 6(b) exhibits Dirac degeneracies at Γ and K, with a corresponding degeneracy at K′ required by time-reversal symmetry. The PC3 unit cell, shown in the inset, consists of six identical dielectric cylinders arranged with sixfold rotational symmetry. The lattice constant is 15 mm, and each cylinder has a radius of 2.7 mm and a relative permittivity of 9.8. The cylinder centers lie 5 mm from the unit-cell center. The bulk eigenstates, band dispersions under PEC truncation, and corresponding CABS field profiles of PC3 are presented in Supplemental Section 8 [49].

To experimentally characterize transmission and channel isolation, we excite the two channels separately. The measured field distributions in Figs. 6(c) and 6(d) show that waves launched into the horizontal PC2 channel and the inclined PC1 channel, respectively, traverse PC3 and continue along their original paths, with little leakage into the intersecting channel. The corresponding simulated field distributions in Figs. 6(e) and 6(f) reproduce the main features of the measured propagation patterns, confirming that the fields remain predominantly confined to the initially excited channel as they pass through the intersection. These results demonstrate that CABS transport in distinct momentum regions can be extended from parallel arrays to intersecting waveguide geometries. The proposed crossing allows two independently addressable channels to share the same physical region without additional isolation regions, vertical separation, or magnetic bias, offering a route toward compact two-dimensional topological photonic circuits with low crosstalk.

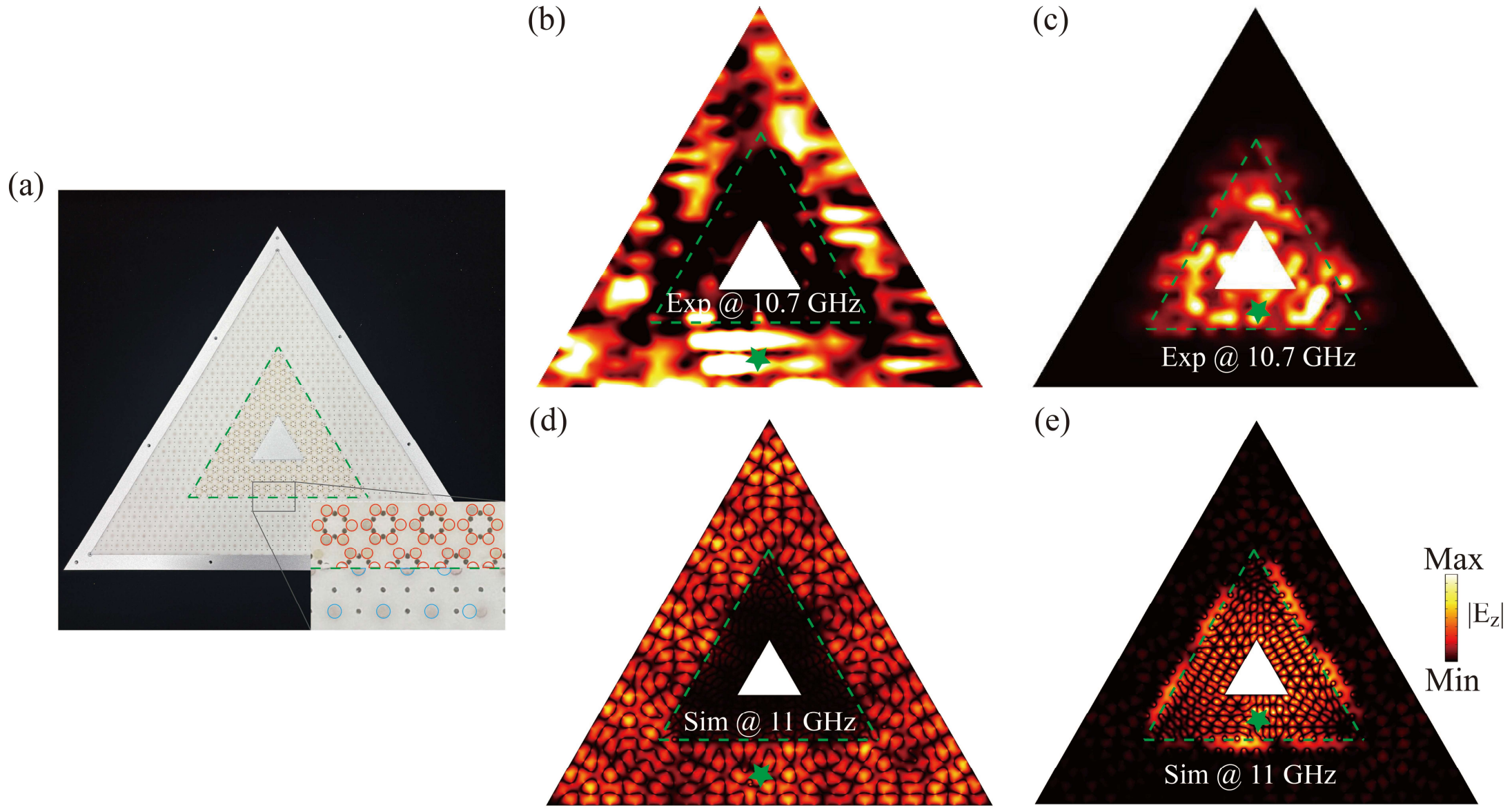

FIG. 5. Insulator-free and low-crosstalk triangular resonator. (a) Photograph of the fabricated sample, consisting of inner (PC1) and outer (PC2) triangular regions. The inset shows a magnified view of the interface (green dashed triangle) between the two regions. (b) and (c) Measured electric-field distributions of the cladding-free triangular resonator modes in the outer (b) and inner (c) regions, respectively. (d) and (e) Simulated electric-field distributions of the cladding-free triangular resonator modes in the outer (d) and inner (e) regions, respectively. The green star indicates the point source.

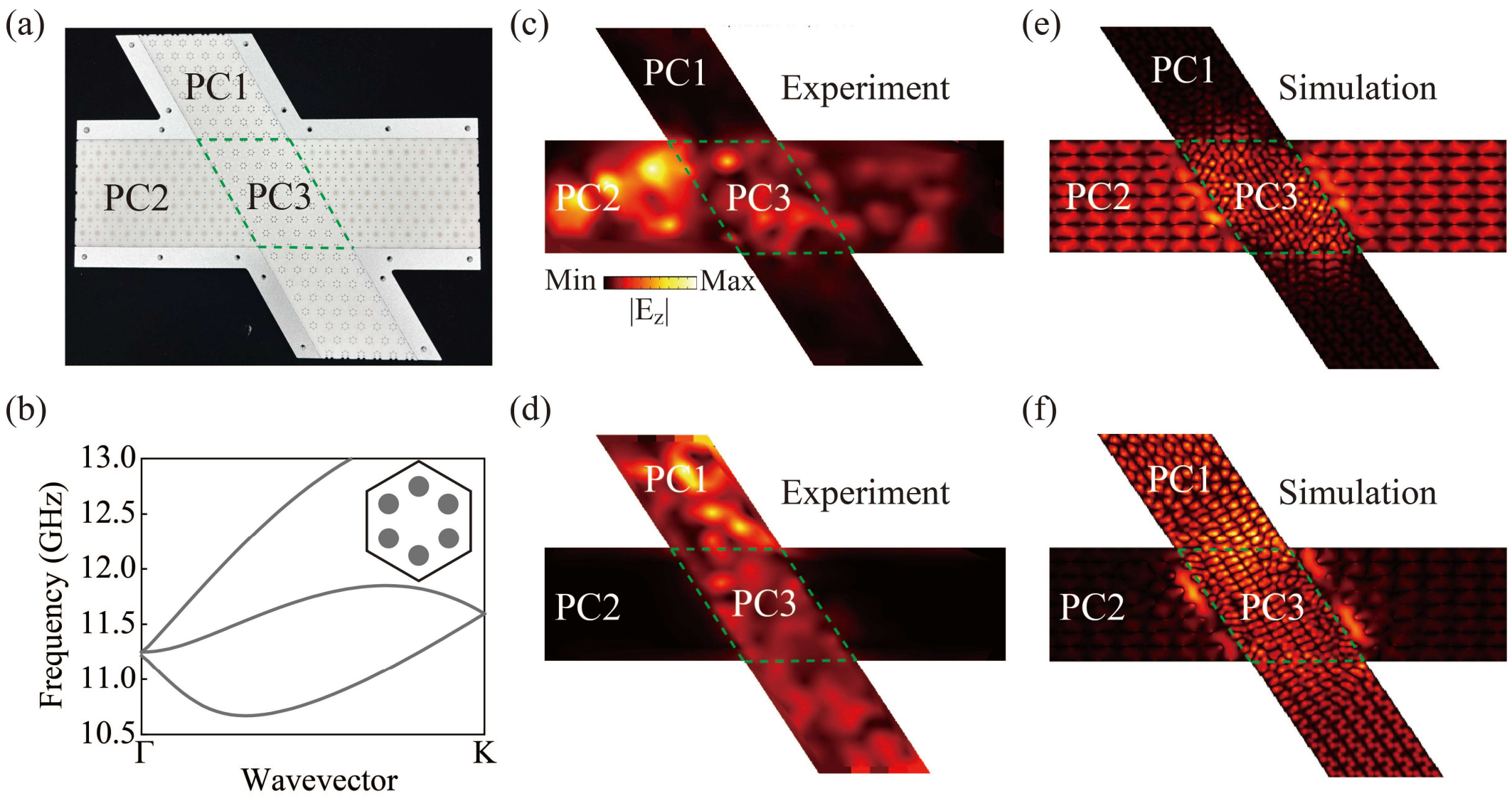

FIG. 6. Low-crosstalk waveguide crossing enabled by momentum-separated CABSs. (a) Photograph of the fabricated waveguide crossing, comprising an inclined PC1 channel and a horizontal PC2 channel. The green dashed parallelogram outlines the PC3 intersection region shared by the two channels. (b) Calculated band structure of PC3, showing Dirac degeneracies at Γ and K that provide momentum-compatible states for the incident CABSs. The corresponding degeneracy at K′ follows from time-reversal symmetry. Inset: unit cell of PC3, consisting of six identical dielectric cylinders arranged with sixfold rotational symmetry. (c) and (d) Measured electric-field distributions under excitation of the horizontal PC2 channel and the inclined PC1 channel, respectively. (e) and (f) Corresponding simulated electric-field distributions.